\documentclass[11pt,prd,preprintnumbers,eqsecnum,superscriptaddress,nofootinbib]{revtex4}
\usepackage[usenames,dvipsnames]{xcolor}
\usepackage{graphics}
\usepackage{graphicx}
\def\be{\begin{equation}}
\def\ee{\end{equation}}
\def\ba{\begin{array}}
\def\ea{\end{array}}
\def\bea{\begin{eqnarray}}
\def\eea{\end{eqnarray}}

\begin{document}

\begin{center}
{\Large \bf $a_{1}$ meson-nucleon coupling constant at finite temperature from the soft-wall AdS/QCD model }

\vskip 1. cm
  {Shahin Mamedov $^{a,b,c}$\footnote{corresponding author: sh.mamedov62@gmail.com }} and
  {Shahnaz Taghiyeva $^{b}$\footnote{ shahnaz.ilqarzadeh.92@mail.ru}},

\vskip 0.5cm

{\it $^a\,$ Institute for Physical Problems, Baku State University,

    Z. Khalilov street 23, Baku, AZ-1148, Azerbaijan,}
    \\ \it \indent $^b$Theoretical Physics Department, Physics Faculty, Baku State University,

     Z.Khalilov street 23, Baku, AZ-1148, Azerbaijan,
\\ \it \indent $^c$Institute of Physics, Azerbaijan National Academy of Sciences,

 H. Javid avenue 33, Baku, AZ-1143
Azerbaijan.\\
\end{center}

\centerline{\bf Abstract} \vskip 4mm
We study the temperature dependence of the $a_1$ meson-nucleon coupling constant in the framework of the soft-wall  AdS/QCD model with thermal dilaton field. Profile functions for the axial-vector and fermion fields in the AdS-Schwarzschild metric are presented. It is constructed an interaction Lagrangian for the fermion-axial-vector-thermal dilaton fields system in the bulk of space-time. From this Lagrangian integral representation for the $g_{a_1NN}$ coupling constant is derived. The temperature dependence of this coupling constant is numerically analyzed.

\vspace{1cm}

\section{Introduction}
The AdS/CFT correspondence is one of the most important discoveries in modern theoretical physics. This principle states the equivalence of two different physical theories, a gravitational theory in the bulk of AdS space-time and a quantum field theory on the boundary of this space-time. Moreover, the duality between these theories is the strong-weak duality, which means that it relates the strong coupling sector of the second theory to the weak coupling sector of the first one \cite{1,2,3,4,5}. Further, this duality was adapted to describe the low-energy dynamics of QCD and called  AdS/QCD \cite{5,6,7,8,9,10}.
There are two approaches in AdS/QCD: top-down and bottom-up ones. The bottom-up approach is based on the direct application of the holographic duality, while the top-down approach to based on the duality between the open and closed string amplitudes. The models of the bottom-up and top-down approaches are widely applied to predict phenomenological quantities in particle physics. These models have a use for solving the problems of the nuclear medium and particle interactions in it. Holographic soft-wall model \cite{7,10,11,12,13}, which is one of two main models in the bottom-up approach, also is valuable for the studies of hadron interactions in the nuclear medium\footnote{more precisely, in the nucleon medium}. This medium is formed as a result of the heavy-ion or proton collisions up to energies when the confinement-deconfinement phase transition occurs. Temperature dependencies of the quantities describing the hadrons in the nuclear medium are investigated in the framework of different approaches and models. The holographic soft-wall model was applied for solving different kinds of problems in the medium: determining of the temperature influence on in-medium quantities such as screening mass \cite{14},  hadron form factors  \cite{15,15,16,17,18,18,20}, quarkonium \cite{21}, decay constants \cite{22}, transport coefficients \cite{23}, anisotropy in  such medium \cite{24} and so on. In the holographic models, the temperature of the medium is taken into account utilizing the black hole temperature in the dual AdS-Schwarzschild background \cite{25,26}. It applied different modifications of the original holographic thermal models to solve phenomenological problems. The thermal sot-wall model was modified in the approach developed in the Refs. \cite{15,16,17,27}, by considering the dilaton field, which is responsible for the chiral and conformal symmetries breaking in the model, as a thermal one, since this field is in the thermal medium. As this field is related to the chiral condensate, the thermalization of the dilaton field in such a way physically means that the authors take into account the temperature dependence of the chiral quark condensate in the medium in the dilaton field as well.  As a check of this idea, the authors of Refs. \cite{15,16,17} investigated the hadron form factors, transition, and electromagnetic form factors of nucleons in the framework of this model and found valid results for them. Continuing these investigations in the Ref. 
\cite{28} the authors have considered the temperature dependence of the vector meson - nucleon coupling constant. It was observed that the value of this constant decreases on temperature increasing and vanishes at the temperature close to the confinement-deconfinement phase transition temperature.  It is interesting to check whether this situation takes place for the other meson sectors of the model or does not. Here we aim to consider the lightest axial-vector meson, namely the $a_1$ meson, and study the temperature dependence of the $a_1$ meson-nucleon coupling constant. 

The paper is organized as follows: In Sec. 2 we briefly review the basic definitions in the soft-wall model having thermal dilaton field and present profile function for the thermal $a_{1}$ meson in this model. In Secs. 3 and 4 the temperature-dependent profile functions for the nucleon and scalar field are briefly derived. In Sec. 5 the integral expressions for the $g_{a_1NN}$  coupling and  $F^2_{a_1}$ decay constants are obtained. In Sec. 6 the temperature dependence of these constants is numerically analyzed. In the conclusion section, we discuss the obtained results.

\section{Soft-Wall Model at Finite Temperature}
 According to AdS/CFT correspondence the quantum field theory at finite temperature on the boundary of the space-time is described by the AdS-Schwarzschild background in the bulk, which has following metric:
\begin{equation}
ds^{2}=g_{MN}dx^{M}dx^{N}=e^{2A(z)}\left(f(z)dt^{2}-dx_{i}dx^{i}-\frac{dz^{2}}{f(z)}\right),
\label{2.1}
\end{equation}
where the $A(z)$ and $f(z)$ functions have explicit form:
\begin{eqnarray}
A(z)=log(R/z),  
\label{2.2}
\end{eqnarray}
\begin{equation}
f(z)=1-\frac{z^{4}}{z_{h}^{4}}. 
\label{2.3}
\end{equation}
  $R$ is the AdS radius and $x=(t,\overrightarrow{x})$ are the set of Minkowski coordinates. The $z$ coordinate is chosen in the  $0\leq z\leq z_{h}$ interval in order to describe the confinement phase of the medium. The black hole Hawking temperature $T$ is related by the position of the black hole horizon $z_h$:
\begin{equation}
T=\frac{1}{4\pi}|\frac{df}{dz}|_{z=z_{h}}=\frac{1}{\pi z_{h}}.
\label{2.4}
\end{equation}
This temperature in the dual QCD theory corresponds to the temperature of hadronic matter.

The main idea of this model is to consider the dilaton field $\varphi(x)$ as one depending on the temperature of the medium \cite{15,16,17}. To this end, the $z$ coordinate was replaced by the Regge-Wheeler tortoise coordinate $r$, which was introduced in Ref. \cite{29}. The relation between these coordinates in the finite temperature limit is following one:
\begin{equation}
r=z[1+\frac{t^{4}}{5}+\frac{t^{8}}{9}+O(t^{12})],
\label{2.5}
\end{equation}
where $t=z/z_h$. An explicit form of the dilaton field is found from the idea of sameness of the temperature dependencies of the dilaton field and quark condensate $\Sigma(T)$. The last one is known from chiral perturbation theory, and we have the dilaton field of the form\footnote{We shall give more explanation in sec. IV.}:
\begin{equation}
\varphi(r,T)=K^{2}(T)r^{2}=(1+\rho(T))k^{2}r^{2}.  \label{2.6}
\end{equation}
The thermal  addition term $\rho_{T}$ has a form:
\begin{equation}
\rho(T)=\delta_{T_{1}}\frac{T^{2}}{12F^{2}}+\delta_{T_{2}}\left(\frac{T^{2}}{12F^{2}}\right)^{2}+O(T^{6}), 
\label{2.7}
\end{equation}
where the constant parameters $\delta_{T_{1,2}}$ denote 
\begin{equation}
\delta_{T_{1}}=-\frac{N_{f}^{2}-1}{N_{f}},\label{2.8}
\end{equation}
\begin{equation}
\delta_{T_{2}}=-\frac{N_{f}^{2}-1}{2N_{f}^{2}}.  \label{2.9}
\end{equation}
Here $N_{f}$ is the number of quark flavors. $F$ is the decay constant in the chiral limit at zero temperature. In Refs. \cite{16,17}  authors find that there is a relation between the soft-wall AdS/QCD model dilaton parameter $k$ and the pion decay constant $F$ in the chiral limit  
\begin{equation}
F=k\frac{\sqrt{3}}{8},  \label{2.10}
\end{equation}
which is right at zero and finite temperatures.
\subsection{Profile function for $a_{1}$ meson at finite temperature}
Let us at first briefly present  the $a_1$ meson at zero temperature following to Ref. \cite{8}. The gauge field sector of the model consists of the  $A_{L,R}$ gauge fields, which are coming from the $SU(2)_{L, R}$ symmetries of the $SU(2)_{L}\times SU(2)_{R}$ flavor symmetry group of the model. $V=\frac{1}{2}(A_{L}+A_{R})$ vector  and $A=\frac{1}{2}(A_{L}-A_{R})$ axial-vector fields are composed of these gauge fields, and the bulk action for the gauge sector can be written in terms of the composite fields:
\begin{eqnarray}
S=-\frac{1}{4g_{5}^{2}}\int_{0}^{\infty} d^{5}x e^{-\varphi(z)}\sqrt{g}Tr[F_{L}^{2}+F^{2}_{R})]=-\frac{1}{2g_{5}^{2}}\int_{0}^{\infty}d^{5}x e^{-\varphi(z)}\sqrt{g}Tr[F_{V}^{2}+F^{2}_{A}].  \label{2.11}
\end{eqnarray}
Here $g$ is $g=det g_{MN}$,  $\varphi(z)=-k^2z^2$ is the dilaton field, $X$ is scalar field. $F_{V,A}$ are the field strength tensors of the vector and axial-vector fields. Here we shall deal with the only axial-vector field, the strength tensor of which has a form:
\begin{equation}
F_{A}^{MN}=\partial^{M}A^{N}-\partial^{N}A^{M}-\frac{i}{\sqrt{2}}[A^{M},A^{N}].  \label{2.12}
\end{equation}
For further calculations the axial gauge $A_5=0$ is chosen. Fourier components $A_{\mu}(q,z)$ of the axial-vector field  satisfies the boundary conditions $A(q,\epsilon)=1$, $\partial_{_{z}}A(q,z=z_{m})=0$ at ultraviolet and infrared boundaries, respectively. Transverse part  $(\partial^{\mu}A_{\mu}=0)$ of the axial-vector field will be decomposed into the Kaluza-Klein modes $A_{\mu}(q,z)=\sum_{n=0}A_{n\mu}(q)A_{n}(z)$ and the equation for this part has the normalizable solutions for the discrete values of the $4D$ momentum $q^{2}=m_{n}^{2}$. Equation of motion leads to the following equation for the $A_{n}(z)$ mode profile function  \cite{8}:
\begin{equation}
\partial_{z}\left(e^{-B(z)}\partial_{z}A_{n}(z)\right)+\left(m_{n}^{2}-g_5^2e^{2A(z)}v^2(z)\right)e^{-B(z)}A_{n}(z)=0. \label{2.13}
\end{equation}
This equation cannot be solved analytically. However, its solution near the UV boundary $(z\rightarrow 0)$ can be found for the $m_q=0$ (chiral limit) case. In these limits, the equation (\ref{2.13}) gets a form, which coincides with the one for the vector field:
\begin{equation}
\partial_{z}\left(e^{-B(z)}\partial_{z}A_{n}\right)+m_{n}^{2}e^{-B(z)}A_{n}=0 \label{2.14}
\end{equation}
and is solved similarly to the vector field case. 
 Making the $A_{n}(z)=e^{B(z)/2}\psi_{n}(z)$ substitution in the equation of motion in (\ref{2.14}), this equation will get the  Schr\"odinger-type equation form and has a solution expressed in the terms of Laguerre polynomials $L^{n}_{m}$  \cite{13}:
 \begin{equation}
 \psi_{n}(z)=e^{-k^{2}z^{2}/2}(kz)^{m+\frac{1}{2}}\sqrt{\frac{2n!}{(m+n)!}}L_{n}^{m}(k^{2}z^{2}). \label{2.15}
 \end{equation}
 Since the UV boundary value of the $A_{\mu}(q,z)$ function corresponds to the wave function of the axial-vector meson on this boundary, we may accept that the UV asymptotic solution (\ref{2.15}) is the wave function of the $a_{1}$ meson. For the $a_{1}$ meson, which is the lightest axial-vector meson, we take $m=1$ in this solution:
 \begin{equation}
 A_{n}(z)=k^{2}z^{2}\sqrt{\frac{2}{n+1}}L_{n}^{1}(k^{2}z^{2}).  \label{2.16}
 \end{equation}
 
As the zero-temperature equation (\ref{2.14}) for the axial-vector meson in the aforementioned limits has the same form as one for the vector field, the thermalization procedure with the thermal dilaton field for the vector field in the Ref. \cite{15} is applicable for the axial-vector field case as well.
Now let us briefly present formulas from this thermalization of vector field applied in Ref. \cite{15}.
In the profile function for the vector field $\Phi_n (r,T)$  a substitution $\phi_{n}(r,T)=e^{-\frac{B_{T}(r)}{2}}\Phi_n (r,T)$ with $B_{T}(r)=\varphi(r,T)-A(r)$ is useful and in the rest frame the equation for $\phi_{n}(r,T)$ obtains a form of Schr\"odinger equation: 
\begin{equation}
\left[-\frac{d^{2}}{dr^{2}}+U(r,T)\right]\phi_{n}(r,T)=M_{n}^{2}(T)\phi_{n}(r,T).
\label{2.17}
\end{equation}
Here $U(r,T)$ is the effective potential, which is written in the sum of the  temperature-dependent and zero-temperature terms:
\begin{equation}
U(r,T) =U(r) + \Delta U(r,T).
\label{2.18}
\end{equation}
 The $U(r)$ and $\Delta U(r,T)$ potentials were found in the following:
\begin{eqnarray}
U(r)=k^{4}r^{2}+\frac{(4m^{2}-1)}{4r^{2}},\nonumber \\
\Delta U(r,T)=2\rho(T)k^{4}r^{2}.
\label{2.19}
\end{eqnarray}
Here $m=N+L-2$. $N$ is number of partons in the meson and $N=2$ for our case. $L$ is the angular momentum and $L=1$ for our case. The meson mass spectrum $M_{n}^{2}$ in (\ref{2.17}) is written in the sum of discrete zero-temperature part $ M_{n}^{2}(0)$ and continuous finite-temperature part $\Delta M_{n}^{2}(T)$:
\begin{eqnarray}
M_{n}^{2}(T) =\ M_{n}^{2}(0)+\Delta M_{n}^{2}(T), \nonumber \\
M_{n}^{2}(0)=4k^2\left(n+\frac{m+1}{2}\right),\nonumber \\
\Delta M_{n}^{2}(T)=\rho(T)M_{n}^{2}(0) + \frac{R\pi^{4}T^{4}}{k^{2}},\nonumber \\
R =(6n-1)(m+1).
\label{2.20}
\end{eqnarray}
Finally, the solution of equation (\ref{2.17}) for the bulk profile $\phi_{n}(r,T)$ was found in the following form \cite{15}:
\begin{equation}
\phi_{n}(r,T)=\sqrt{\frac{2\Gamma(n+1)}{\Gamma(n+m+1)}}K^{m+1}r^{m+\frac{1}{2}}e^{-\frac{K^{2}r^{2}}{2}}L_{n}^{m}(K^{2}r^{2}),
\label{2.21}
\end{equation}
which coincides with the zero-temperature solution found in Ref. \cite{13} on replacing $r\rightarrow z$, $K(T) \rightarrow k$.

 Thus, in the chiral limit and near the UV boundary, as the finite-temperature profile function of the axial-vector field can be taken the (\ref{2.21}) solution, and the thermal $A_{n}(z, T)$ in $r$ coordinate will have the form below:
\begin{equation}
A_{n}(r,T)=K^{2}z^{2}\sqrt{\frac{2}{n+1}}L_{n}^{1}(K^{2}r^{2}).   \label{2.22}
\end{equation}
\section{NUCLEON PROFILE FUNCTION AT FINITE TEMPERATURE}
In the bottom-up approach of the holographic QCD to describe the left and right-handed components of the nucleons, two fermion fields $\left(N_1, N_2\right)$ are introduced in the bulk of AdS space, which are not interrelated \cite{30,31,32,33}. Following Ref. \cite{16}, here we present briefly the solution of the equation of motion for the fermion fields interacting with the thermal dilaton field. Action for such fermion field $N(x,r,T)$  in the background (\ref{2.1}) and in terms of the  $r$ coordinate will be written as 
\begin{equation}
S=\int d^{4}x dre^{-\varphi(r,T)}\sqrt{g}{\bar{N}}(x,r,T)D_{\pm }(r)N(x,r,T),
\label{3.1}
\end{equation}
where the $D_{\pm }(r)$ covariant derivative contains a temperature-dependent part as well:
\begin{equation}
D_{\pm}(r) =\frac{i}{2}\Gamma^{M}\left[\partial_{M}-\frac{1}{4}\omega_{M}^{ab}\ \left[\Gamma_{a}\Gamma_{b}\right]\right]\mp\left[\mu_F(r, T)+U_{F}(r, T)\right].
\label{3.2}
\end{equation}
The "mass" $\mu_F (r, T)$ of the $N(x, r, T)$ thermal fermion field is related to the  $\mu_F$ mass at zero-temperature:
\begin{equation}
\mu_F(r, T)=\mu_F\ f^{\frac{3}{10}}(r, T),
\label{3.3}
\end{equation}
and $\mu_F$ defined in the following equation 
\begin{equation}
\mu_F=\emph N_{B}+\emph L-\frac{3}{2}.
\label{3.4}
\end{equation}
Here $\emph N_{B}=3$ is the number of partons in the composite fermion corresponding to the nucleon, and $\emph L$ is the orbital angular momentum. The temperature-dependent potential  $U_{F}(r, T)$ for the fermion field is related to the dilaton field and the  $f(r, T)$ blackening function:
\begin{equation}
U_{F}(r, T)=\varphi (r,T)/f^{\frac{3}{10}}(r, T).
\label{3.5}
\end{equation}
Non-zero components of the spin connection $ \omega_{M}^{ab} $ are given by equation:
\begin{equation}
\omega_{M}^{ab}=(\delta_{M}^{a}\delta_{r}^{b}-\delta_{M}^{b}\delta_{r}^{a})\ r f^{\frac{1}{5}}(r, T). 
\label{3.6}
\end{equation}
The $\sigma^{MN}=\left[\Gamma^{M},\Gamma^N\right]$ in the (\ref{3.2}) derivative is the commutator of the Dirac $\Gamma^{M}$ matrices in the curved space-time, and these matrices are related to the reference frame $\Gamma^{a}$ matrices by the $\Gamma^{M} =e_{a}^{M}\Gamma^{a}$ relation. Inverse vielbeins  are $e_{a}^{M}=r\times diag \{\frac{1}{f(r)},1,1,1,-f(r)\} $. Reference frame $\Gamma^{a}$ matrices are chosen as $ \Gamma^{a}=(\gamma^{\mu },\ -i\gamma ^{5}) $.      
For the fifth component of the fermion field we choose an axial gauge $N_{5}(x,r,T)=0$ and decompose $N(x,r,T)$ into the left- and right-chirality components:
\begin{equation}
N(x,r,T)=N^{R}(x,r,T)+N^{L}(x,r,T),
\label{3.7}
\end{equation}
where the chiral components are defined as $ N^{R}(x,r,T)= \frac{1-\gamma^{5}}{2}N(x,r,T)$,   $N^{L}(x,r,T)=\frac{1+\gamma^{5}}{2}N(x,r,T)$.

Kaluza-Klein decomposition for these components is written in terms of the profile functions $\Phi_{n}^{L,R}(r,T)$, which are temperature-dependent as well:
\begin{equation}
N^{L,R}(x,r,T)=\sum_{n} N_{n}^{L,R}(x)\Phi_{n}^{L,R}(r,T).
\label{3.8}
\end{equation}
For the nucleons we consider the $L=0$ case and the total angular momentum will be  $\emph{J}=\frac{1}{2}$. It is useful to write the  $\Phi_{n}^{L,R}(r,T)$ profiles with the prefactors $e^{-\frac{3}{2}A(r)}$:
\begin{equation}
\Phi_{n}^{L,R}(r,T)=e^{-\frac{3}{2}A(r)} F_{n}^{L,R}(r,T).
\label{3.9}   
\end{equation}
After the substitution of these profile functions into the equations of motion in the rest frame of nucleon $(\vec{p}=0)$, the following equations for the $F_{n}^{L, R}(r, T)$ profile functions were obtained in  the Ref. \cite{16}:
\begin{equation} 
\left[\partial_{r}^2+U_{L,R}(r,T)\right]F_{n}^{L,R}(r,T)=M_{n}^{2}(T)F_{n}^{L,R}(r,T).
\label{3.10}
\end{equation}
The spectrum $M_{n}^{2}(T) $ is divided into the temperature-dependent part, which is continuous, and the "cold" part, which is quantized. The quantized part 
\begin{equation}
M_{n}^{2}(T)=4K^2(T)\left(n+m+\frac{1}{2}\right)=4k^2\left(1+\rho (T)\right)\left(n+m+\frac{1}{2}\right)
\label{3.11}
\end{equation}
at $T=0$ coincides with the known zero-temperature spectrum \cite{8}. The effective potentials $U_{L,R}(r,T)$ in Eq. (\ref{3.10}) are written in the sum of zero- and finite-temperature  parts:
\begin{eqnarray}
U_{L,R}(r,T) =U_{L,R}(r)+\Delta U_{L,R}(r,T),
\nonumber \\
\Delta U_{L,R}\left(r,T\right) =2\rho (T)k^{2}\left(k^{2}r^{2}+m \mp \frac{1}{2}\right).
\label{3.12}
\end{eqnarray}
Here
\begin{equation}
m=N_B+L-\frac{3}{2}.
\label{3.13}
\end{equation}
Finally, solutions to the equations (\ref{3.10}), which are profile functions for the boundary thermal nucleons, were found in the form \cite{16}:
\begin{eqnarray}
F_{n}^{L}(r,T)=\sqrt{\frac{2\Gamma (n+1)}{\Gamma (n+m_{L}+1)}}K^{m_{L}+1}r^{m_{L}+\frac{1}{2}}e^{-\frac{K^{2}r^{2}}{2}}L_{n}^{m_{L}}\left(K^{2}r^{2}\right), \nonumber\\
F_{n}^{R}(r,T)=\sqrt{\frac{2\Gamma (n+1)}{\Gamma (n+m_{R}+1)}}K^{m_{R}+1}r^{m_{R}+\frac{1}{2}}e^{-\frac{K^{2}r^{2}}{2}}L_{n}^{m_{R}}\left(K^{2}r^{2}\right),
\label{3.14}
\end{eqnarray}
where
$m_{L,R}=m\pm\frac{1}{2}$.
The $\Phi_{n}(r,T)$ and $F_{n}(r,T)$  profile functions obey normalization conditions:
\begin{equation}
\int_{0}^{\infty}dr e^{-\frac{3}{2}A(r)}\Phi_{m}^{L,R}(r,T)\Phi_{n}^{L,R}(r,T))=\int_{0}^{\infty }dr F_{m}^{L,R}(r,T) F_{n}^{L,R}(r,T)=\delta_{mn}. 
\label{3.15} 
\end{equation} 
With the replacements $r\rightarrow z$ and $K(T) \rightarrow k$ the  (\ref{3.14}) functions coincide with the profiles at zero temperature case (see Ref. \cite{13}). 

\section{Bulk Vacuum expectation value of scalar meson field}
Action for the pseudo-scalar $X$ field in AdS/QCD has a form:
\begin{equation}
	S=\int_{0}^{\infty}d^{5}x\sqrt{g}e^{-\varphi(z)}Tr\{|DX|^{2}-m_{5}^{2}|X|^{2}\}.  \label{4.1}
\end{equation}
In terms of the $r$ tortoise coordinate  and with the thermal dilaton field, this action looks slightly changed:
\begin{equation}
	S_{X}=\int d^{4}xdr\sqrt{g}e^{-\varphi(r,T)}Tr\left[|DX|^{2}+3|X|^{2}\right],
	\label{4.2}
\end{equation}
where $DX$ is the covariant derivative including the interaction with the gauge fields $A_{L,R}$. In terms of  the vector $M_M$ and  the axial-vector $A_M$ fields terms this derivative will be written as: $D^{M}X=\partial^{M}X-iA_{L}^{M}X+iXA_{R}^{M}=\partial^{M}X-i\left[M_M,X\right]-i\{A_M,X\}$, $A_{L,R}^{M}=A_{L,R}^{M}t^{a}$, and $F_{L,R}^{MN}$ are the field strengths of these fields. The $X$ field  transforms under the bifundamental representation of the flavor symmetry group $SU(2)_{L}\times SU(2)_{R}$ of the model and performs the breaking of this symmetry by Higgs mechanism \cite{30,31,32,34,35}. E. o. m. for this field, which is obtained from the action (\ref{4.1}), has a solution: 
\begin{equation}
\langle X \rangle=\frac{1}{2}v(z),  \label{4.3}
\end{equation}
where
\begin{equation}
v(z)=\frac{1}{2}M_{q}az+\frac{1}{2a}\Sigma z^{3},  \label{4.4}
\end{equation}
where $a=\sqrt{N_c}/(2\pi)$  $(N_c=3)$ is the normalization parameter \cite{30,31,32}.According to the bulk/boundary correspondence dictionary, the parameters $M_{q}$ and  $\Sigma$ are identified with the $u, d$ quark mass matrix and with the chiral condensate $\Sigma=<0|\bar{q}q|0> $ correspondingly.
For the finite temperature case, the quark condensate in the solution (\ref{4.4}) depends on temperature and, the $z$ coordinate should be replaced by the tortoise one in it. Then, in the solution (\ref{4.4}) the replacement $\Sigma\rightarrow \Sigma(T)$ will be done and it accepts a form: 
\begin{equation}
v(r,T)=\frac{1}{2}M_{q}ar+\frac{1}{2a}\Sigma(T) r^{3}.  \label{4.5}
\end{equation}
In the Ref. \cite{36} an explicit form of the temperature dependence of the quark condensate $\Sigma(T)$ was defined using two-loop chiral perturbation theory at finite temperature in the following:
\begin{equation}
\Sigma(T)=\Sigma[1-\frac{N_{f}^{2}-1}{N_{f}}\frac{T^{2}}{12F^{2}}-\frac{N_{f}^{2}-1}{2N_{f}^{2}}(\frac{T^{2}}{12F^{2}})^{2}+O(T^{6})]=\Sigma[1+\Delta_{T}+O(T^{6})],  \label{4.6}
\end{equation}
where  $N_{f}$ is the number of quark flavor and $F$ is the decay constant. 
The dilaton field  $\phi(r)$ is responsible for the dynamical breaking of the chiral symmetry in AdS/QCD. The chiral condensate formed as a result of this symmetry breaking in the boundary QCD.  Both constants have dependence only on temperature, so, in Refs. \cite{15,16,17} authors  supposed that the temperature dependence of the $\Sigma(T)$ quark condensate should be the same as the temperature dependence of the dilaton parameter  $K^2(T)$:
\begin{equation}
K^2 (T)=k^2\frac{\Sigma(T)}{\Sigma}.
\label{4.7}
\end{equation}
In addition, it was conjectured that known zero-temperature relation between the quantities quark condensate  $\Sigma$, number of flavors  $N_f$, condensate parameter $B$  and the  pseudo-scalar meson decay constant $F$ in the chiral limit
\begin{equation}
\Sigma=-N_{f}BF^{2}
\end{equation}
holds for the finite temperature case as well:
\begin{equation}
\Sigma(T)= - N_{f}B(T)F^{2}(T).
\label{4.8}
\end{equation}
Then, according to the (\ref{2.6}) and (\ref{4.7}) relations the $\Sigma(T)$ dependence will be written in the following  form \cite{35}:
\begin{equation}
\Sigma(T)=\Sigma\left[1+\rho(T)\right].
\label{4.9}
\end{equation}
Let us note, that the relation (\ref{4.10}) is correct up to $T^6$ degree of the temperature. The $ F(T)$ and  $B(T)$ dependencies  have been studied in the Ref. \cite{36}.
\section{$g_{a_1NN}$ coupling constant}
 Action for the interaction  between the axial-vector, fermion and scalar fields in the soft-wall model with the thermal dilaton will be written employing interaction Lagrangian $L(x,z,T)$:
\begin{equation}
S=\int_{0}^{\infty}d^{4}xdz\sqrt{g}e^{-\varphi(z,T)}L(x,z,T). \label{5.1}
\end{equation}
 In terms of the $r$ tortoise coordinate, this action will be written without the exponent factor:
\begin{equation}
S=\int_{0}^{\infty}d^{4}xdr\sqrt{g}L(x,r,T). \label{5.2}
\end{equation}
 The meson-nucleon coupling constant, which we want to investigate here, can be derived from the bulk interaction action involving the gauge, bulk scalar and fermion fields. Let us list here possible interactions at the lowest order of the fields. Corresponding Lagrangian terms should be Hermitian scalars under the existing symmetries in the model, namely gauge, chiral and 5D "Lorentz" symmetries. The minimal gauge coupling term in 5D Lagrangian is usual one \cite{13,30,31,32,38,39}:
\begin{equation}
L^{(0)}=\frac{1}{2}[\overline{\Psi}_{1} \Gamma^{M}A_{M} \Psi_{1}-\overline{\Psi}_{2}\Gamma^{M}A_{M}\Psi_{2}].   \label{5.3}
\end{equation}
Another well-known interaction term, which obeys the symmetries mentioned above, is the bulk magnetic gauge coupling \cite{27,28,12,25,33,34}:
\begin{equation}
L^{(1)}=\frac{i}{2}k_{1}[\overline{\Psi}_{1}\Gamma^{MN}F_{MN}\Psi_{1}+\overline{\Psi}_{2}\Gamma^{MN}F_{MN}\Psi_{2}].    \label{5.4}
\end{equation}
Triparticle interaction of the bulk spinors, gauge  and scalar fields was constructed in the Ref.  \cite{38,40} and is similar to Yukawa coupling term introduced in \cite{30,31,32}:
\begin{equation}
L^{(2)}=g_{Y}[\overline{\Psi}_{1}X\Gamma^{M}A_{M} \Psi_{2}+\overline{\Psi}_{2}X^{\dagger} \Gamma^{M}A_{M} \Psi_{1}].    \label{5.5}
\end{equation}
According to AdS/CFT correspondence, the generating functional in the gauge field theory is equivalent to the exponential of an on-shell action in the gravity theory (is called GKP-W relation) \cite{21,22,23}:
\begin{equation}
Z_{AdS}[\phi_{0}]=\langle e^{i\int dx\phi_{0}(x)O(x)}\rangle_{gauge}=e^{iS_{gravity}[\phi_{0}]}    \label{5.6}
\end{equation}
where $Z_{AdS}[\phi_{0}]$ is the generating functional with the source $\phi_{0}$ coupled with an operator $O(x)$ and $S_{gravity}$ is an on-shell action with boundary condition $\phi\rightarrow\phi_{0}$ at the UV boundary. In our case, the axial-vector current $(J_{\mu})$ is an operator and its holographic dual is the axial-vector field $A_{\mu}$. The holographic principle will give us the following relation between these quantities:
\begin{equation}
\langle J_{\mu}\rangle=-i\frac{\delta Z_{QCD}}{\delta A_{\mu}^{0}}|_{A_{\mu}^{0}=0}.    \label{5.7}
\end{equation}
Here $Z_{QCD}=e^{iS_{int}}$ is the generating functional for the boundary QCD, $A_{\mu}^{0}=A_{\mu}(q,z=0)=A_{\mu}(q)$ is a boundary value of the axial-vector field.
In the boundary QCD theory, the four-dimensional  axial-vector current of nucleons is defined as follows:
\begin{equation}
J_{\mu}(p^{\prime},p)=g\overline{u}(p^{\prime})\gamma^{5}\gamma_{\mu}u(p).   \label{5.8}
\end{equation}
For the nucleon-$ a_1 $ meson interaction the $ g $ constant is the $g_{a_{1}NN}$ coupling constant and it can be determined from the equivalence of the right-hand sides of Eqs. (\ref{5.7}) and (\ref{5.8}). After integrating out the space-time coordinates from the 5D action integral, we obtain integral expressions for the $g_i(r, T)$ constants corresponding to the  $L^{(i)}$ Lagrangian terms:
\begin{equation}
g_0(r,T)=\frac{1}{2}\int^{\infty}_{0}\frac{dr}{r^{4}}A_{0}(r,T)(|\phi_{1R}(r,T)|^{2}-|\phi_{1L}(r,T)|^{2}),    \label{5.9}
\end{equation}
\begin{equation}
g_1(r,T)=\frac{k_{1}}{2}\int^{\infty}_{0}\frac{dr}{r^{3}}\partial_{z}A_{0}(r,T)(|\phi_{1R}(r,T)|^{2}+|\phi_{1L}(r,T)|^{2}), \label{5.10}
\end{equation}
\begin{equation}
g_2(r,T)=2g_{Y}\int^{\infty}_{0}\frac{dr}{r^{4}}A_{0}(r,T)v(r,T)(\phi_{1R}(r,T)\phi_{1L}(r,T)).    \label{5.11}
\end{equation}
Temperature dependence of each $g_i(r,T)$ constant can be studied numerically. In addition, we can investigate the temperature dependence of the $a_1$ meson decay constant, which was defined in the Ref. \cite{8} for the 'cold' case by the  holographic formula:
\begin{equation}
F_{a_1}^2=\frac{1}{g_5^2}\left[v^{\prime\prime}(0)\right]^2. \label{5.12}
\end{equation}
For the finite-temperature case this decay constant will be calculated with the thermal profile $ A_n(r,T) $ in Eq. (\ref{2.8}):
\begin{equation}
F_{a_1}^2\left(T\right)=\frac{1}{g_5^2}\left[A^{\prime\prime}_n(r,T)|_{r=0} \right]^2 \label{5.13}
\end{equation}
and the temperature dependence of this constant also can be investigated numerically.
\section{Numerical analysis}
 To visualize the temperature dependence of the terms in the Eqs. (\ref{5.9}), (\ref{5.10}), (\ref{5.11}) and the constant in the Eq. (\ref{5.13}) we have performed numerically an integration over the $r$ variable of these temperature-dependent integrals.The values of parameters were given in GeV units. We use the values of the parameters $k_{1}$, $m_{q}$, $\Sigma$ and $g_{Y}$, which were fixed in the earlier works. The $k_{1}=-0.98$ value of the parameter was obtained from the fitting of the couplings $g_{\pi NN}=13.5$ and $g_{\rho NN}=-8.6$ of the ground state nucleons in hard-wall AdS/QCD model in  Refs. \cite{15,36}. The numerical values of the quark condensate and quark mass were taken from Ref. \cite{34} and are $\Sigma=(0.213$ MeV$)^{3}$, $m_{q}=8.3$ MeV,  respectively. The constant $g_{Y}=9.182$ was fixed to get nucleon mass $m_{N}=0.94$  GeV in the framework of hard-wall \cite{30,31,32}. The numerical values of the number of quark flavors $N_f$ and the pseudo-scalar meson decay constant $F$ were taken from the Refs. \cite{15,36}. We consider ground state of the $a_1$ meson and take $n=0$ in the profile function (\ref{2.22}).
  In Figs. 1-4 we present the numerical results for the temperature dependencies of the $g_0$, $g_1$, $g_2$ terms and for the $g_{a_{1}NN}$ coupling constant. As we seen in the section IV, the dilaton parameter $K^2(T)$ is defined through the quark condensate $\Sigma (T)$, which is determined by the $N_f$ and $F$ parameters. So, in order to analyse how the $g_i$ terms and the $g_{a_{1}NN}$ constant depend on values of the $N_f$ and $F$ parameters, in these figures we have taken following values of these parameters:  $N_f=2$, $F=87$ MeV; $N_f=3$, $F=100$ MeV and $N_f=5$, $F=140$ MeV, which were applied in the Refs. \cite{15,16,17}. In Fig.5 we have plotted the temperature dependence of the squared decay constant  $F_{a_1}^2$ of the $a_1$ meson.
 As is seen from the graphs, all $g(T), F_{a_1}^2(T)$ dependencies decrease and become zero around the temperature value $T=200$ MeV, which is the confinement-deconfinement phase transition temperature $T_c$.

\section{Conclusion}
 The numerical analysis here for the $g_{a_{1}NN}$ coupling constant shows that the temperature dependence of this constant has a similar shape of behavior, as was for the $g_{\rho NN}$ constant, i.e., the $g_{a_{1}NN}$ constant decreases with increasing the temperature and near the $T_c$ temperature it becomes zero. A similar result was obtained in \cite{41}, where authors studied pion-nucleon coupling constant at finite temperature\footnote{We thank the referee for bringing this work to our attention.}.  The shape of dependence almost does not depend on the values of the $N_f$ and $F$ parameters. Also, from a comparison of the graphics we observe, that the minimal coupling is not the leading term in the $a_1$ meson-nucleon interaction. It should note, the vanishing quantities at the $T_c$ temperature were obtained in the works \cite{15,16,17,25} as well, where was applied AdS-Schwarzchild metrics. 

As was noted in the Ref. \cite{28}, the study of couplings near the $T_c$ temperature may have a use for understanding hadron matter forming in the early Universe. For completeness of these investigations, it is reasonable to make a similar analysis in the scalar (pseudo-scalar) meson sector of this model and to check the temperature dependency results, which were obtained here and in the Ref. \cite{28}, for the pion-nucleon coupling constant. This question is under consideration.

\newpage

\begin{figure}[!h]
	\begin{center}
		\includegraphics[width=6cm]{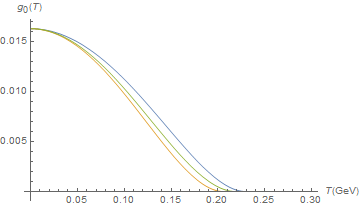}
	\end{center}
	\caption{ $g_0(T)$ term at $N_{f}=2$, $F=87$ MeV (orange line); $N_{f}=3$, $F=100$ MeV (green line); $N_{f}=5$, $F=140$ MeV (blue line) }
	\label{f2MB}
\end{figure}
\begin{figure}[!h]
	\begin{center}
		\includegraphics[width=6cm]{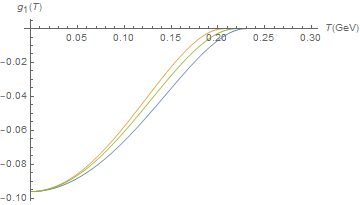}
	\end{center}
	\caption{ $g_1(T)$ term at $N_{f}=2$, $F=87$ MeV (green line); $N_{f}=3$, $F=100$ MeV (blue line); $N_{f}=5$, $F=140$ MeV (orange line)}
	\label{f2MB}
\end{figure}
\begin{figure}[!h]
	\begin{center}
		\includegraphics[width=6cm]{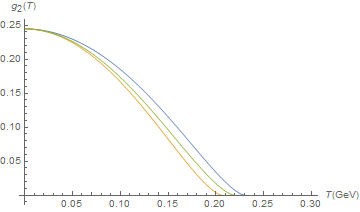}
	\end{center}
	\caption{$g_2(T)$ term at $N_{f}=2$, $F=87$ MeV (green line); $N_{f}=3$, $F=100$ MeV (blue line); $N_{f}=5$, $F=140$ MeV (orange line) }
	\label{f2MB}
\end{figure}
\begin{figure}[!h]
	\begin{center}
		\includegraphics[width=6cm]{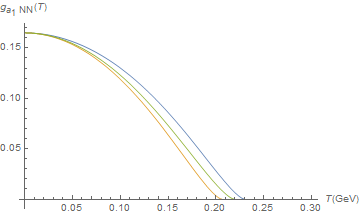}
	\end{center}
	\caption{$g_{a_1NN}(T)$ constant at $N_{f}=2$, $F=87$ MeV (orange line); $N_{f}=3$, $F=100$ MeV (green line); $N_{f}=5$, $F=140$ MeV (blue line) }
	\label{f2MB}
\end{figure}
\begin{figure}[!h]
	\begin{center}
		\includegraphics[width=6cm]{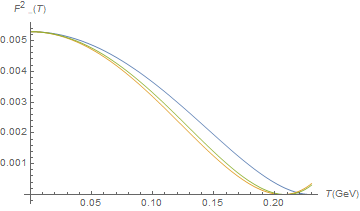}
	\end{center}
	\caption{Temperature dependence of the $F^2_{a_1}(T)$ decay constant at $N_{f}=2$, $F=87$ MeV (orange line); $N_{f}=3$, $F=100$ MeV (green line); $N_{f}=5$, $F=140$ MeV (blue line).}
	\label{f2MB2}
\end{figure}


\begin{thebibliography}{99}
\bibitem{1}
J. M. Maldacena, Adv. Theor. Math. Phys. {\bf{2}}, 231 (1998);
\bibitem{2}
S.S. Gubser, I.R. Klebanov and A.M. Polyakov, Phys. Lett. B {\bf{428}}, 105 (1998) [arxiv:hep-th/9802109];
\bibitem{3}
E. Witten, Adv. Theor. Math. Phys. {\bf{2}}, 253, (1998), [arxiv:hep-th/9802150]; 
\bibitem{4}
E. Witten, Adv. Theor. Math. Phys. {\bf{2}}, 505, (1998), [arxiv:hep-th/9803131]; 
 \bibitem{5}
Horatiu Nastase, Introduction to AdS-CFT [arXiv:0712.0689 [hep-th]];
 \bibitem{6} 
G.F.de Teramond and S.J. Brodsky, Phys. Rev. Lett. {\bf{94}}, 201601 (2005), [arXiv:hep-th/0501022];  
 \bibitem{7}
J. Erlich, E. Katz, D.T. Son and M.A. Stephanov, Phys. Rev. Lett. {\bf{95}}, 261602 (2005), [arXiv:hep-ph/0501128];
 \bibitem{8}
A. Karch, E. Katz, D.T. Son and M.A. Stephanov, Phys. Rev. D {\bf{74}},  (2006) [arxiv:0602229 [hep-ph]];
 \bibitem{9}
H. Boschi-Filho and N.R.F. Braga, J. High Energy Phys. 0305 (2003) 009 [arXiv:0212207[hep-th]];
 \bibitem{10}
H. Boschi-Filho and N.R.F. Braga, Eur. Phys. J. C {\bf{32}}, 529  (2004), [arXiv: 0209080 [hep-th]];
 \bibitem{11}
H.R. Grigoryan and A.V. Radyushkin, Phys. Rev. D {\bf{76}}, 115007 (2007), [arXiv:07090500[hep-ph]];
 \bibitem{12}
Z. Abidin and C. Carlson, Phys. Rev. D {\bf{79}}, 115003 (2009), [arXiv:0903.4818[hep-ph]];
 \bibitem{13}
T. Gutsche, V. E. Lyubovitskij, I. Schmidt, and A. Vega, Phys. Rev. D {\bf{85}}, 076003 (2012) [arXiv: 1108.0346 [hep-ph]];
 \bibitem{14}
X. Cao, S. Qiu, H. Liu and D. Li, J. High Energy Phys. 08 (2021) 005 [arXiv:2102.10946[hep-ph]];
 \bibitem{15}
T. Gutsche, V. E. Lyubovitskij, I. Schmidt, and A. Y. Trifonov, Phys. Rev. D {\bf{99}}, 054030 (2019) [arXiv:1902.01312  [hep-ph]];
 \bibitem{16}
T. Gutsche, V. E. Lyubovitskij, I. Schmidt, and A. Y. Trifonov, Phys. Rev. D {\bf{99}}, 114023 (2019) [arXiv:1905.02577 [hep-ph]];
 \bibitem{17}
T. Gutsche, V. E. Lyubovitskij and I. Schmidt, Nucl. Phys. {\bf{B952}} (2020) 114934 [arXiv:1906.08641 [hep-ph]];
 \bibitem{18}
E.F. Capossoli, M.A. Martin Contreras, D. Li, A. Vega and H. Boschi-Filho, Phys. Rev. D {\bf{102}}, 086004 (2020), [arXiv:2007.09283 [hep-ph]]; 
 \bibitem{19}
M.A. Martin Contreras, E.F. Capossoli, D. Li, A. Vega and H. Boschi-Filho "Proton and neutron form factors from deformed gravity/gauge duality" (2021) [arXiv:2108.05427 [hep-ph]];
 \bibitem{20}
G. Ramalho and D. Melnikov, Phys. Rev. D {\bf{97}}, 034037  (2018);
 \bibitem{21}
M.A. Martin Contreras, S. Diles and A. Vega, Phys. Rev. D {\bf{103}}, 086008 (2021)  [arXiv:2101.06212 [hep-ph]];
 \bibitem{22}
M.A. Martin Contreras and A. Vega Phys. Rev. D {\bf{101}}, 046009 (2020) [arXiv:1910.10922 [hep-th]];
 \bibitem{23}
D. Li, S. He and M. Huang, J. High Energy Phys. 06 (2015) 046 [arXiv:1411.5332  [hep-ph]];
 \bibitem{24}
I. Aref'eva, K. Rannu and P. Slepov,  J High Energy Phys. 06 (2021) 090 [arXiv:2009.05562  [hep-th]];
 \bibitem{25}
P. Colangelo, F. Giannuzzi, S. Nicotri, and V. Tangorra, Eur. Phys. J. C {\bf{72}}, 2096 (2012);
 \bibitem{26}
L.A.H. Mamani, A.S. Miranda, H. Boschi-Filho and N.R.F. Braga, J. High Energy Phys.{\bf{03}}, 058 (2014);
 \bibitem{27}
A. Vega and M. A. Martin Contreras, Nucl.Phys. {\bf{B942}}, 410 (2019) [arXiv:1808.09096 [hep-ph]];
 \bibitem{28} 
Sh. Mamedov and N. Nasibova, Phys. Rev. D {\bf{104}}, 036010 (2021)  [arxiv:2103.10494  [hep-ph]];
 \bibitem{29}
T. Regge and J.A. Wheeler, Phys. Rev. {\bf{108}}, 1063, (1957);
 \bibitem{30}
D.K. Hong, T. Inami and H.-U. Yee, Phys.Lett. B {\bf{646}}, 165 (2007) [arXiv:0609270[hep-ph]]; 
 \bibitem{31}
H.Ch. Ahn, D.K. Hong, Ch. Park and S. Siwach, Phys. Rev. D {\bf{80}}, 054001 (2009), [arxiv:0904.3731[hep-ph]];
 \bibitem{32} 
N. Maru, M. Tachibana, Eur.Phys. J. C  {\bf{63}}, 123, (2009) [arxiv:0904.3816[hep-ph]];
 \bibitem{33}
Z. Abidin and C. Carlson, Phys. Rev. D {\bf{79}}, 115003 (2009),[arXiv:0903.4818[hep-ph]];
 \bibitem{34} 
A. Cherman, Th.D. Cohen and E.S. Werbos, Phys. Rev. C {\bf{79}}, 045203, (2009), [arxiv:0804.1096[hep-ph]];
 \bibitem{35}
J. Chen, S. He, M. Huang and D. Li, J. High Energy Phys. {\bf{01}}, 165 (2019);
 \bibitem{36}
J. Gasser and H. Leutwyler, Phys. Lett. B {\bf{184}} 83 (1987);
 \bibitem{37}
C. P. Herzog, Phys. Rev. Lett. {\bf{98}}, 091601 (2007). 
 \bibitem{38} 
N. Huseynova and Sh. Mamedov, Int.J.Mod.Phys. A {\bf{34}} 35, 1950240  (2019); 
 \bibitem{39}
N. Huseynova and Sh. Mamedov, Int.J.Theor.Phys. {\bf{54}}  10, 3799 (2015);
 \bibitem{40} 
Sh. Mamedov, B.B. Sirvanli, I. Atayev and N. Huseynova, Int. J. Theor. Phys. {\bf{56}}, 1861 (2017) [arxiv:1609.00167v2 [hep-th]];
 \bibitem{41} 
C.A. Dominguez, C. van Gend, M. Loewe, Phys.Lett. B 429 (1998) 64 [arXiv:9803469 [hep-ph]].
 
\end{thebibliography}
\end{document}